\documentclass[aps,pra,twocolumn,amssymb,amsfonts,amsmath,showpacs,final,a4paper,superscriptaddress]{revtex4}

\usepackage{float}
\usepackage{color}
\usepackage[dvips,final]{graphicx}
\newcommand{\be}{\begin{equation}}
\newcommand{\ee}{\end{equation}}
\newcommand{\bea}{\begin{eqnarray}}
\newcommand{\eea}{\end{eqnarray}}
\newcommand{\pd}{\partial}

\begin{document}

\title{Coherent excitations and electron phonon coupling in Ba/EuFe$_2$As$_2$ compounds
investigated by femtosecond time- and angle-resolved photoemission spectroscopy}

%\author{I.~Avigo$^{1}$, R. Cort\'{e}s$^{2,3}$\footnote{current address: Center for Functional Nanomaterials, Brookhaven National Laboratory, Upton New York, 11973, USA}, L. Rettig$^{1,2}$, S. Thirupathaiah$^4$, H.S. Jeevan$^5$, P. Gegenwart$^{5}$, T. Wolf$^6$,
%M. Ligges$^1$, M. Wolf$^3$, J. Fink$^4$ and U. Bovensiepen$^{1}$}

%\email {uwe.bovensiepen@uni-due.de}
%\homepage {www.uni-due.de/agbovensiepen}

%\address{Fakult\"{a}t f\"{u}r Physik, Universit{\"a}t Duisburg-Essen, Lotharstr. 1, D-47048 Duisburg, Germany}
%\address{Fachbereich Physik, Freie Universit{\"a}t Berlin, Arnimallee 14, D-14195 Berlin, Germany}
%\address{Abt. Physikalische Chemie, Fritz-Haber-Institut d. MPG, Faradayweg 4-6, D-14195 Berlin, Germany}
%\address{Leibniz-Institute for Solid State and Materials Research Dresden, P.O.Box 270116, D-01171 Dresden, Germany}
%\address{I. Physik. Institut, Georg-August Universit\"{a}t G\"{o}ttingen, D-37077 G\"{o}ttingen, Germany}
%\address{I. Karlsruhe Institute of Technology, Institut f\"{u}r Festk\"{o}rperphysik, D-76021 Karlsruhe, Germany}

\author{I. Avigo}
\affiliation {Fakult\"{a}t f\"{u}r Physik, Universit{\"a}t Duisburg-Essen, Lotharstr. 1, D-47048 Duisburg, Germany}

\author{R. Cort\'{e}s}
%\affiliation{\textcolor{red}{current address:} Center for Functional Nanomaterials, Brookhaven National Laboratory, Upton New York, 11973, USA}
\affiliation{Fachbereich Physik, Freie Universit{\"a}t Berlin, Arnimallee 14, D-14195 Berlin, Germany}
\affiliation{Abt. Physikalische Chemie, Fritz-Haber-Institut d. MPG, Faradayweg 4-6, D-14195 Berlin, Germany}

\author{L. Rettig}
\affiliation {Fakult\"{a}t f\"{u}r Physik, Universit{\"a}t Duisburg-Essen, Lotharstr. 1, D-47048 Duisburg, Germany}
\affiliation {Fachbereich Physik, Freie Universit{\"a}t Berlin, Arnimallee 14, D-14195 Berlin, Germany}

\author{S. Thirupathaiah}
\affiliation{Leibniz-Institute for Solid State and Materials Research Dresden, P.O.Box 270116, D-01171 Dresden, Germany}

\author{H.S. Jeevan}
\affiliation{I. Physik. Institut, Georg-August Universit\"{a}t G\"{o}ttingen, D-37077 G\"{o}ttingen, Germany}

\author{P. Gegenwart}
\affiliation{I. Physik. Institut, Georg-August Universit\"{a}t G\"{o}ttingen, D-37077 G\"{o}ttingen, Germany}

\author{T. Wolf}
\affiliation{Karlsruhe Institute of Technology, Institut f\"{u}r Festk\"{o}rperphysik, D-76021 Karlsruhe, Germany}

\author{M. Ligges}
\affiliation {Fakult\"{a}t f\"{u}r Physik, Universit{\"a}t Duisburg-Essen, Lotharstr. 1, D-47048 Duisburg, Germany}

\author{M. Wolf}
\affiliation{Abt. Physikalische Chemie, Fritz-Haber-Institut d. MPG, Faradayweg 4-6, D-14195 Berlin, Germany}

\author{J. Fink}
\affiliation{Leibniz-Institute for Solid State and Materials Research Dresden, P.O.Box 270116, D-01171 Dresden, Germany}

\author{U. Bovensiepen}
\email{uwe.bovensiepen@uni-due.de}
\homepage {www.uni-due.de/agbovensiepen}
\affiliation {Fakult\"{a}t f\"{u}r Physik, Universit{\"a}t Duisburg-Essen, Lotharstr. 1, D-47048 Duisburg, Germany}

\date{\today}

%##################################################################################################
\begin{abstract}

\footnotesize{We employed femtosecond time- and angle-resolved photoelectron
spectroscopy to analyze the response of the electronic structure
of the 122 Fe-pnictide parent compounds Ba / EuFe$_2$As$_2$ and
optimally doped BaFe$_{1.85}$Co$_{0.15}$As$_2$ near the $\Gamma$
point to the optical excitation by an infrared femtosecond laser
pulse. The experiments were carried out at an equilibrium temperature of 100
and $300\,\mathrm{K}$} with a pump fluence below $1.4\,\mathrm{mJ/cm^2}$.
We identify pronounced changes of the electron population within several $100\,\mathrm{meV}$ above and
below the Fermi level, which we explain as combination of (i)
coherent lattice vibrations, (ii) a hot electron and hole
distribution, and (iii) transient modifications of the chemical
potential. The response of the three different materials is very
similar. In the Fourier transformation of the time-dependent
photoemission intensity we identify three modes at 5.6, 3.3, and
$2.6\,\mathrm{THz}$. While the highest frequency mode is safely assigned to
the A$_{1g}$ mode, the other two modes require a discussion in
comparison to literature. The time-dependent evolution of
the hot electron distribution follows a simplified description of
a transient three temperature model which considers two heat baths
of lattice vibrations, which are more weakly and strongly coupled to transiently excited electron population.
Still the energy transfer from electrons to the strongly coupled
phonons results in a rather weak, momentum-averaged
electron-phonon coupling quantified by values for $\lambda
\langle\omega^2\rangle$ between 30 and $70\,\mathrm{meV^2}$. The
chemical potential is found to present pronounced transient changes reaching a
maximum of $15\,\mathrm{meV}$ about $0.6\,\mathrm{ps}$ after optical excitation and 
a damped oscillatory behavior resembling the intensity
oscillations assigned to coherent phonons. This change in the
chemical potential is particularly strong in a two band
system like in the 122 Fe-pnictide compounds investigated here
due to the pronounced variation of the electrons density of
states close to the equilibrium chemical potential.
\end{abstract}
\maketitle

%#################################################################################################
\section{Introduction}

The discovery of high-$T_c$ superconductivity in F doped LaFeAsO
~\cite{Kamihara2008} led to an almost immediate  discovery of
further FeAs-based superconductors with transition temperatures
$T_c$ up to $55\,\mathrm{K}$. While the superconducting pairing mechanism in
conventional superconductors is related to electron-phonon
coupling, current thinking is that the iron pnictides (FePn),
similar to the cuprate superconductors, belong to the class of
unconventional superconductors, in which the mechanism is possibly
related to correlation effects and/or magnetic
excitations~\cite{Stewart2011}. Although there are various
similarities between cuprates and FePns, e.g. the nearness in the
phase diagram of antiferromagnetism and superconductivity, there
are also differences. While in the cuprates the parent compounds
are antiferromagnetic Mott-Hubbard insulators, in the FePns the
parent compounds are antiferromagnetic metals which indicates that
correlation effects in the latter systems are reduced. This view
is supported by e.g. x-ray absorption spectroscopy on
FePns resulting in a moderate value of the on-site Coulomb repulsion of $U
\approx1.5\,\mathrm{eV}$ \cite{Kroll2008}.

On the other hand there is no real consensus on the mechanism of
high-$T_c$ superconductivity in FePns and on the role of
electron-phonon coupling. Experiments on the isotope
effect in these compounds exhibit conflicting
results~\cite{Stewart2011}. Theoretical calculations on the
electron-phonon coupling constant $\lambda$ yielded for the
non-magnetic and the antiferromagnetic system $\lambda\approx$ 0.2
and $\lambda <$ 0.35, respectively~\cite{Boeri1(2008), Boeri2010}.
Optical pump-probe studies with femtosecond time-resolution
analyzed $\lambda$ by determining the second moment of the
Eliashberg electron-phonon coupling function
$\lambda\langle\omega^2\rangle$ through the rate of energy
transfer from the optically excited electrons to phonons
~\cite{Mansart2009, Stojchevska2010}. Although the reported results
vary in detail due to different assumptions regarding $\omega$ the
resulting electron-phonon coupling constant $\lambda$ is found to
be 0.1 -- 0.2. In our recent femtosecond time- and angle-resolved
photoemission spectroscopy (trARPES) study~\cite{Rettig2012}, in
which ultrashort light pulses in the UV spectral range were used,
we derived for the second moment of the Eliashberg electron-phonon
coupling function, $\lambda\langle\omega^2\rangle = 90 \pm 40\,\mathrm{meV^2}$.
Using an average phonon frequency $\langle\omega\rangle
\approx20\,\mathrm{meV}/\hbar$ a value of $\lambda <$ 0.2 can be derived
from this result. Although these electron-phonon coupling
constants are small, which makes it difficult to explain $T_c$
values above $20\,\mathrm{K}$ by electron-phonon coupling, there is no
consensus on the role of phonons in the mechanism of
superconductivity in FePns. Therefore information on phonons
and their coupling to the charge carriers is  vitally important.
In this view we present here further results on the coherent
excitations of phonons and on the electron-phonon coupling in
undoped Ba/EuFe$_2$As$_2$ and and Co doped Ba$_2$As$_2$ 122 compounds using trARPES.

Photoelectron spectroscopy is widely considered to be a surface
sensitive method due to the small penetration depth of the free
electron like final state into the material for energies of 10 --
$100\,\mathrm{eV}$. For the low photon energy of 6 -- $7\,\mathrm{eV}$ used in laser
photoemission experiments~\cite{Kiss2005,Koralek2006,Rettig2012}
as well as in the present study a bulk sensitivity can be
considered~\cite{Kiss2005} due to a larger mean free path of
excited electrons. However, on the one hand, a sensitivity to
electronic surface states for these small kinetic energies of
photoelectrons is well established~\cite{loukakos_PRL07,bovensiepen10}.
On the other hand, combined
surface and bulk sensitivity~\cite{Perfetti2006} as well as
detection of states at buried interfaces~\cite{Rettig2012a} were
reported. Therefore, the degree of surface vs. bulk sensitivity
might be highly material specific and the photoemission matrix
elements need to be considered in detail~\cite{Schattke_2000}. The
present study investigates 122 compounds with their pronounced two-dimensional
electronic structure. On the other hand 122 compounds exhibit an increasing
three-dimesionality with increasing doping~\cite{Thirupathaiah2010}.
As will be presented below, the reported
phenomena do not depend on doping. Therefore, the question of
surface vs. bulk sensitivity appears not to be an essential one in
the present context and for the following we consider to probe
with trARPES the surface near region.

%##################################################################################################

\section{\label{exp}Experimental details}

trARPES experiments were performed on EuFe$_2$As$_2$,
BaFe$_2$As$_2$ parent compounds and an optimally doped BaFe$_{1.85}$Co$_{0.15}$As$_2$ ($T_c=23\,\mathrm{K}$). Single crystals
of EuFe$_2$As$_2$ were grown by the Bridgman method~\cite{Jeevan2008} while Ba-based compounds were grown from
self-flux in an alumina crucible~\cite{Hardy(2009)}.
All samples were cleaved in ultrahigh vacuum ($p<10^{-10}\,\mathrm{mbar}$) at $T=100\,\mathrm{K}$ where the major part of the measurements were carried out.
Additional measurements on the parent compounds were performed at $300\,\mathrm{K}$.

A commercial regenerative amplifier Coherent RegA 9050 generates ultrashort infrared laser
pulses of $820\,\mathrm{nm}$ wavelength ($1.5\,\mathrm{eV}$) with a repetition rate of $300\,\mathrm{kHz}$ and pulse duration of $55\,\mathrm{fs}$. The beam is then split in two parts, where one is used to provide the pump pulses and the other one is frequency-doubled and compressed twice to achieve $205\,\mathrm{nm}$ wavelength ($6\,\mathrm{eV}$) pulses with a duration
of $80\,\mathrm{fs}$, used for probing. We detect photoelectrons which are generated by the UV probe pulse using an electron time of flight spectrometer with an
acceptance angle of $\pm3$ degrees. Thereby we monitor the momentum and energy dependent single-particle spectral function in the vicinity of the Fermi level, $E_F$, as function of pump-probe delay. Measurements were carried out in normal emission, i.e. around the $\Gamma$-point. The energy resolution of $50\,\mathrm{meV}$ is determined by the time of flight spectrometer and the bandwidth of the probe pulses. The overall temporal resolution was $100\,\mathrm{fs}$~\cite{Rettig2012}.\\

%##################################################################################################

\section{\label{Results}Results and discussion}

\begin{figure}[tb]
    \resizebox{8.5cm}{!}{\includegraphics{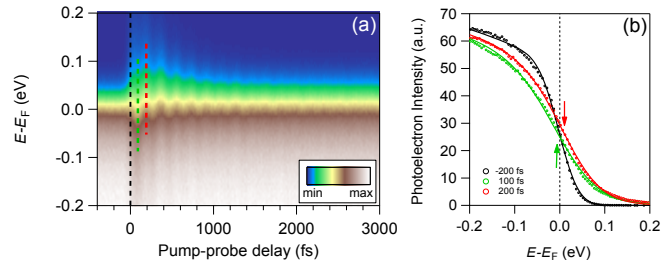}}%.eps
    \caption{\label{fig:f1_spectrum+Ef}\footnotesize{
    (a) Color coded intensity plot of the trARPES intensity of BaFe$_{1.85}$Co$_{0.15}$As$_2$ at the $\Gamma$ point as a function of binding energy and
    pump-probe delay, taken with an incident pump fluence of $F=1.4$ mJ/cm$^2$ at $T=100$ K. Red and green dashed lines mark the spectra shown in (b).
    (b) trARPES spectra for the first minimum (red) and maximum (green) of the oscillation in comparison to a spectrum before excitation (black).
    The solid lines are fits to the data (see section \ref{Osc_Ef}). The Fermi level position $E_F$ extracted from the fit is indicated by the red and
    green arrows.}}
\end{figure}

The time-resolved photoelectron intensity of BaFe$_{1.85}$Co$_{0.15}$As$_2$ at the $\Gamma$-point is shown in
figure \ref{fig:f1_spectrum+Ef}(a) as function of binding energy and time delay for an incident excitation fluence of
$F=1.4\,\mathrm{mJ/cm^2}$. After optical excitation, (i) pronounced oscillations modulate the
photoemission spectrum around $E_F$ and (ii) optically excited electrons and holes broaden the distribution function around $E_F$. In this report we disentangle
these contributions and discuss the coherent phonon and electron dynamics in sections \ref{Coherent_modes} and \ref{Osc_Ef}, respectively.
Figure \ref{fig:f1_spectrum+Ef}(b) compares spectra taken at $\Delta t = 100\,\mathrm{fs}$ and $\Delta t = 200\,\mathrm{fs}$, corresponding to the first minimum and maximum of
the oscillations, with a spectrum taken before excitation. Apart from the strong change of the distribution function at $E_F$ after excitation, a rigid shift of
the cutoff at $E_F$ induced by the oscillation is observed.

% FIGURE 1: Spectrum

%##################################################################################################

\subsection{\label{Coherent_modes}Coherent phonon dynamics}

\begin{figure}[tb]
    \resizebox{8.5cm}{!}{\includegraphics{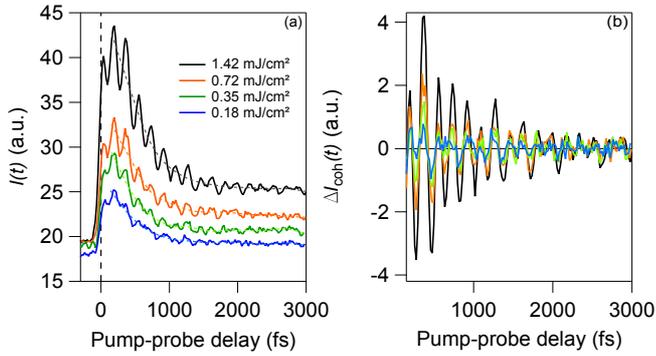}}%.eps
    \caption{\label{fig:f2 spline}\footnotesize{
    (a) Integrated trARPES intensity of BaFe$_{1.85}$Co$_{0.15}$As$_2$ as a function of pump-probe delay for four different fluences measured at 100 K.
    Dashed lines are the fitted spline function used to obtain background subtraction. In panel (b) the respective background-subtracted oscillations are shown.}}
\end{figure}

\begin{figure*}[tb]
    \resizebox{.9\textwidth}{!}{\includegraphics{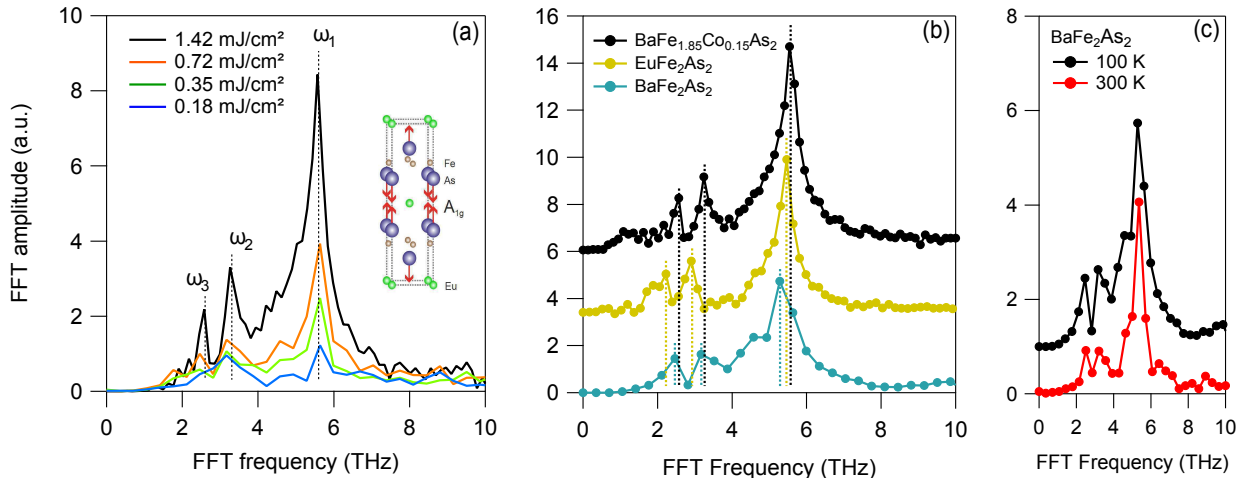}}%.eps
    \caption{\label{fig:f3 FFT}\footnotesize{
    (a) Fast Fourier amplitudes of the data in figure \ref{fig:f2 spline} for correspondent fluences showing the three coherent modes $\omega_1$=5.6 THz,
    $\omega_2$=3.3 THz, $\omega_3$=2.6 THz. The inset is a sketch of the A$_1g$ mode, corresponding to a vertical displacement of the As atoms with
    respect to the Fe plane. (b) Comparison of FFTs for BaFe$_{1.85}$Co$_{0.15}$As$_2$, BaFe$_2$As$_2$ and EuFe$_2$As$_2$ at $T=100$ K. Traces are vertically offset for clarity. (c) Comparison of FFTs of BaFe$_2$As$_2$ for $T$=300 K and $T$=100 K. All three modes are found to be independent of $T$.}}
\end{figure*}

We start with the discussion of the pronounced oscillations recognized in figure \ref{fig:f1_spectrum+Ef}(a).
The time-dependent photoemission intensity integrated for energies $E>E_F$, $I (t)$, is depicted in figure
\ref{fig:f2 spline}(a) for different pumping fluences. This integration yields a rather good signal-to noise ratio due to the high contrast at the Fermi cutoff.
Within the first picosecond after optical excitation, an oscillation with a period close to $200\,\mathrm{fs}$ is observed. For $\Delta t$ $>$ $1.5\,\mathrm{ps}$, an interference pattern is found which indicates the presence of further modes. This is particularly evident for the data at higher fluence.
These oscillations are superimposed on a steep rise in integrated intensity after the optical excitation followed by an
intensity decrease. In the following we refer to the oscillatory part as the coherent contribution and to the non-oscillatory part as the incoherent
contribution to $I(t)$. To determine the coherent contribution, the incoherent contribution was subtracted from $I(t)$. As a first step,
the incoherent part was described by a single exponential decay. While this description was adequate at low excitation densities, for higher
excitation densities this procedure does not provide a satisfying background determination. For this reason, we adopted a method consisting of fitting a smooth
spline function to the data, which was originally developed for the analysis of extended x-ray absorption fine structure (EXAFS) \cite{Lee81}
and was recently used to extract the coherent oscillations of the charge density wave (CDW) amplitude mode in TbTe$_3$~\cite{Schmitt(2011)}.
This method provides a smooth background while leaving the oscillation unaffected as shown in figure \ref{fig:f2 spline}(b). For more details
see~\cite{Schmitt(2011)}.
After background substraction, fast Fourier transformation (FFT) was used to obtain the frequency spectrum of the coherent contribution
$\Delta I_{coh}(t)$. Figure \ref{fig:f3 FFT}(a) depicts the FFTs for the corresponding fluences presenting a sharp intense peak at
$\omega_{1}=5.6\,\mathrm{THz}$ ($23\pm1\,\mathrm{meV}$) and two weaker modes at $\omega_{2}=3.3\,\mathrm{THz}$ ($14\pm1\,\mathrm{meV}$) and $\omega_{3}=2.6\,\mathrm{THz}$ ($11\pm\,\mathrm{meV}$).
As discussed below these features are likely to be fingerprints of coherent lattice vibrations and are observed in trARPES due
to the coupling between the electronic system and the coherent phonons. All three modes were found with similar frequencies also in the parent
compounds BaFe$_2$As$_2$ and EuFe$_2$As$_2$ (figure \ref{fig:f3 FFT}(b)) suggesting that these coherent excitations are a general phenomenon in 122 FePns.

% FIGURE 2 and 3:

\begin{figure}[tb]
   \resizebox{6.5 cm}{!}{\includegraphics{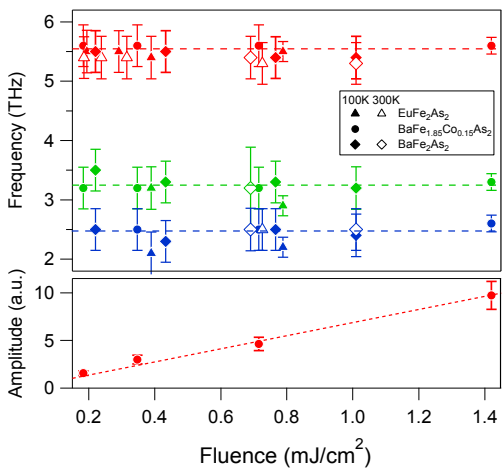}}%.eps
   \caption{\label{fig:f4 Ampl vs Flu}\footnotesize{
    Upper panel: FFT frequencies of the three modes as function of excitation fluence $F$ for EuFe$_2$As$_2$ (triangles),
    BaFe$_{1.85}$Co$_{0.15}$As$_2$ (circles) and BaFe$_2$As$_2$ (diamonds). Open and closed symbols represent data taken at $T=300\,\mathrm{K}$ and $T=100\,\mathrm{K}$,
    respectively. Error bars are defined by the inter-point spacing of the FFT. Horizontal lines are guide to the eyes. Within error bars, the frequencies of the modes do not depend on the excitation density. Lower panel: FFT amplitude for $\omega_1$ of BaFe$_{1.85}$Co$_{0.15}$As$_2$ as function of incident pumping fluence.
    The dashed line is a linear fit to the data.}}
\end{figure}

The resulting frequencies are summarized in figure \ref{fig:f4 Ampl vs Flu} as a function of the incident pumping fluence for the three samples
under investigation. Open and closed symbols denote data taken at $T=300\,\mathrm{K}$ and $T=100\,\mathrm{K}$, respectively.
The experimental errors are determined by the point spacing of the FFT which depends on the extend of the respective data set in the time domain. Additional
uncertainties arise from the decay of the oscillations in the time domain, that leads to the broadening of peaks in the FFT. We find that the determined
FFT frequencies do not vary with the excitation density in all three cases and are temperature-independent (see also figure \ref{fig:f3 FFT}(c)). The FFT amplitude for $\omega_{1}$ in BaFe$_{1.85}$Co$_{0.15}$As$_2$ in the lower panel of figure \ref{fig:f4 Ampl vs Flu} is well described by a
linear dependence with fluence. In combination with the fluence independent frequencies, this points to
the linear regime of the coherent excitations, showing no indication of anharmonicity within our sensitivity~\cite{Hase(02)}.

%FIGURE 4

To determine the geometry and character of the coherent modes, we compare our data with Raman scattering results~\cite{Chauvi`ere2009, Chauvi`ere2011, Choi2008, Kum11, Litvinchuk, Sug11} and a theoretical study~\cite{Boeri1(2008)}. By this comparison, we clearly identify the mode $\omega_1$ at $23\,\mathrm{meV}$ with the Raman active As A$_{1g}$ phonon mode, corresponding to a displacement of the As atoms perpendicular to the FeAs layers, as sketched in the inset of figure \ref{fig:f3 FFT}(a). This mode was found in Raman scattering studies of various pnictide 122 compounds~\cite{Chauvi`ere2009, Chauvi`ere2011, Choi2008, Kum11, Litvinchuk, Sug11},
including BaFe$_{2}$As$_{2}$. In addition, coherent excitation of this mode has also been observed in time-resolved reflectivity measurements on
Co doped BaFe$_{2}$As$_{2}$~\cite{Mansart2009} and in terahertz spectroscopy on BaFe$_{2}$As$_{2}$~\cite{kim2012}. The frequency was found to be $5.56\,\mathrm{THz}$ in the first case and $5.5\,\mathrm{THz}$ in the latter, which is in agreement with our observation.
A more recent study found a coherent excitation of the A$_{1g}$ mode also in SmFeAsO~\cite{Mer10}, which suggests that
the coherent excitation of this mode is a general phenomenon in the FePns. However, the other two modes were not observed with time-resolved
optical spectroscopy.

For the excitation of coherent phonons, traditionally two models have been established~\cite{Garrett1996, Hase(05), Ish10},
the displacive excitation of coherent phonons (DECP)~\cite{Zeiger(DECP)} and the impulsive stimulated Raman scattering (ISRS)~\cite{Yan85}.
While the DECP mechanism is supposed to be the main excitation mechanism in opaque media like metals and semi-metals and
allows to first order only the excitation of fully symmetric A$_1$ modes~\cite{Zeiger(DECP)}, the ISRS is commonly accepted as the excitation process
in transparent media like insulators and some semiconductors and allows for the excitation of all Raman active modes~\cite{Hase(05), Mansart2009}.
It was shown that ISRS can also be extended to opaque materials and is the more general description~\cite{Garrett1996}.
From the observation of coherently excited phonons here in trARPES, we can conclude on a strong coupling of the coherent
vibration to electronic states directly at $E_F$, as we measure the imprint of the coherent modes on the electronic system.

The assignment of the other two modes is less clear. A possible candidate for the $\omega_2$ mode could be the E$_g$ mode, a shear vibration of the
Fe and As atoms within the FeAs plane~\cite{Litvinchuk}. On the one hand this Raman active mode was found at $16\,\mathrm{meV}$~\cite{Litvinchuk} corresponding
to $3.8\,\mathrm{THz}$, which is considerably higher than the mode $\omega_2$ in our experiment and this frequency for the E$_g$ mode is consistent with Raman
observations of this mode in BaFe$_2$As$_2$~\cite{Chauvi`ere2009}. On the other hand, a Raman scattering study on CaFe$_2$As$_2$~\cite{Litvinchuk2011} reported the
observation of a mode at $3.35\,\mathrm{THz}$, which could be attributed to the E$_g$ mode. In addition, a recent study of Co doped BaFe$_2$As$_2$ using
inelastic x-ray scattering found modes close to the BZ center with an energy $\approx$ $13\,\mathrm{meV}$~\cite{Reznik2009}. These values are
all consistent with our observation.

Comparing our frequencies with the calculated phonon dispersion in~\cite{Boeri1(2008)}, we see that another possible
candidate for the $\omega_2$ mode could be the mode at the X-point around $13\,\mathrm{meV}$, which shows combined in-plane and out-of-plane oscillations with
wave vector $q$ equal to the nesting wave vector $Q_n$ and enhanced electron-phonon (e-ph) coupling according to the calculations. The coherent excitation of phonon modes with finite momenta is usually prohibited, as the exciting photons allow only vertical transitions due to their negligible momentum. However,
a specific momentum dependent optical excitation probability could in principle generate a transient momentum dependent population of hot charge
carriers that could drive also bosons modes with finite momenta. Furthermore, contributions of surface phonons or spin excitations~\cite{Mel08, Inosov2010},
which do not show up in the calculations or in Raman spectroscopy, cannot be ruled out.

Regarding the $\omega_3$ mode, the calculation in~\cite{Boeri1(2008)}
shows a mode at the zone-center around $10\,\mathrm{meV}$ which is, however, not Raman active. Furthermore, according to the calculation,
this mode couples only weakly to the electronic system and thus would not be observed here. Therefore additional work is required
in order to unambiguously clarify the geometry and character of these two modes, $\omega_2$ and $\omega_3$.\\

% FIGURE 5:
\begin{figure}[tb]
    \resizebox{8 cm}{!}{\includegraphics{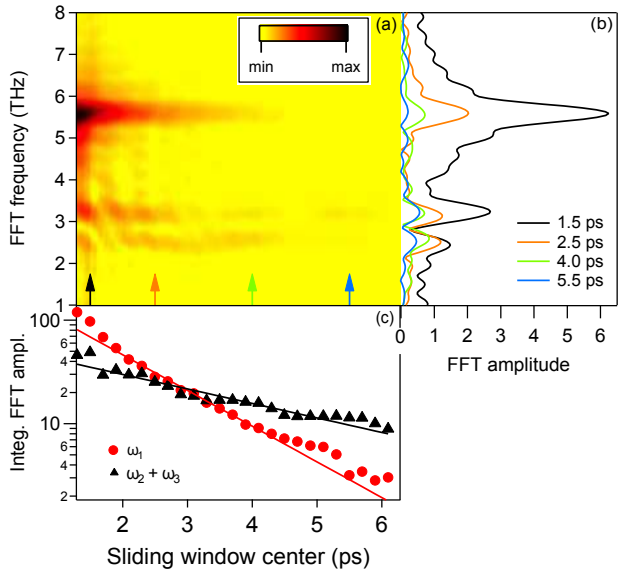}}%.eps
    \caption{\label{fig:f5 SWFT}\footnotesize{
    (a) Color coded intensity plot of the swFFT intensity of BaFe$_{1.85}$Co$_{0.15}$As$_2$ (at $F=1.4$ mJ/cm$^2$, $T= 100$ K) as function of frequency and time delay of the sliding window center. A sliding window width of $2.4\,\mathrm{ps}$ was used. (b) swFFT corresponding to four different sliding window centers, marked
    by arrows in (a). (c) Time evolution of the integrated swFFT amplitude of $\omega_1$ and $\omega_2$+$\omega_3$ integrated together, plotted
    on a logarithmic scale. Solid lines are fits to the data.}}
\end{figure}

We will now concentrate on the time evolution of the three modes. Considering the FFTs in figure \ref{fig:f3 FFT}, we
notice that the mode $\omega_1$ exhibits a considerably broader peak in the frequency domain than the other two modes $\omega_2$ and
$\omega_3$. This indicates a faster relaxation in the time domain of this mode. In order to study this time evolution in more detail, we
concentrated on the BaFe$_{1.85}$Co$_{0.15}$As$_2$ data using the highest pump fluence of $F=1.4\,\mathrm{mJ/cm^2}$ and applied a sliding window fast
Fourier transform (swFFT) analysis, with a time-window of $2.4\,\mathrm{ps}$ which allows a frequency
resolution of $\sim 0.5\,\mathrm{THz}$, sufficient to resolve the two modes $\omega_2$ and $\omega_3$ that are closest to each other. The intensity of the swFFT
signal as function of the transformed frequency and the sliding window center is shown in figure \ref{fig:f5 SWFT}(a). No change in the frequency of the modes
with time is observed, which is consistent with the harmonic regime of the oscillations concluded above and from figure \ref{fig:f4 Ampl vs Flu}. In addition, the amplitude of
$\omega_{1}$ is indeed damped with a higher decay rate than the amplitude of the other two modes. This becomes obvious in figure \ref{fig:f5 SWFT}(b), where
the swFFT intensity corresponding to four different sliding window centers, marked by arrows in figure \ref{fig:f5 SWFT}(a), is shown. The integrated swFFT amplitude at
the frequencies of the modes is plotted on a logarithmic scale as a function of the window position in figure \ref{fig:f5 SWFT}(c). Note that due to interference
effects between $\omega_2$ and $\omega_3$, combined amplitudes corresponding to these two modes were integrated. The integrated swFFT
amplitude of both $\omega_1$ and $\omega_2+\omega_3$ shows a clear linear behavior in the logarithmic plot. By fitting single exponential decay functions to
the respective amplitude time evolution (solid lines), we determine $\tau_1=1.2\pm0.1\,\mathrm{ps}$ and $\tau_{\omega_2+\omega_3}=3.1\pm0.4\,\mathrm{ps}$ as decay times.
We will now discuss the different relaxation times for the mode $\omega_1$ and the modes $\omega_2$ and $\omega_3$. The relaxation of coherently
excited phonons occurs both through pure dephasing and population decay of the excited modes. While the pure dephasing takes place mainly by scattering at
crystal defects and e-ph interaction, the population decay is governed by anharmonic phonon-phonon (ph-ph) interaction~\cite{MHase(2010)}. The
scattering with crystal defects is unlikely to explain the different decay behavior of different modes within the same crystal, assuming a
homogeneous excitation density of these modes. This could be different for e.g. surface modes. The other two relaxation processes can be distinguished
by their temperature and fluence dependence. On the one hand, the relaxation by anharmonic ph-ph decay into two acoustic phonons shows an increase in
decay rate with increasing temperature~\cite{Hase(05), MHase(2010)}. This is not observed in our case, as we have not found an increase of the width
of the peaks in the FFT spectra for the higher temperature as shown in figure \ref{fig:f3 FFT}(c). On the other hand, the scattering of coherent phonons with excited carriers by e-ph coupling is highly dependent on the density of excited carriers and thus on the excitation density, which leads to a faster decay at
higher excitation densities~\cite{MHase(2010), Hase(02)}. This is indeed observed in the data in figure \ref{fig:f3 FFT}(a), where we find an increase with fluence of the linewidth of $\omega_1$.
Therefore, we conclude that the dominating relaxation channel for the coherent oscillations is the coupling to excited carriers by e-ph scattering. The
shorter lifetime of the mode $\omega_1$ compared to the other two modes thus indicates a stronger coupling to the electronic system for the mode
$\omega_1$. This is consistent with the lower amplitude of the other two modes, as e-ph coupling is the key parameter for both the generation and the
detection of coherent phonons in trARPES and, as a consequence, the amplitude of coherent phonons in the detected
signal scales with the coupling strength.

Interestingly, the relaxation that we have found for the mode $\omega_1$ is clearly faster than in the time-resolved optical reflectivity measurements,
where the A$_{1g}$ mode was found to relax with a time-constant $\tau_{A_{1g}}\sim 2.5\,\mathrm{ps}$~\cite{Mansart2009}. This might be explained by a
difference in crystal quality, exhibiting less crystal defects in the optical pump-probe experiments. Another possibility might be the different
detection technique used. Whereas the optical reflectivity probes the entire region of the optical penetration depth of several tens of nanometers,
photoelectron spectroscopy is a surface sensitive technique. Thus, the coherent phonons in the sub-surface region might exhibit a faster relaxation
than in the bulk of the lattice.

%##################################################################################################

\subsection{Dynamics of the electronic system}
\label{Osc_Ef}

%FIGURE 6
\begin{figure}[tb]
    \resizebox{7.5cm}{!}{\includegraphics{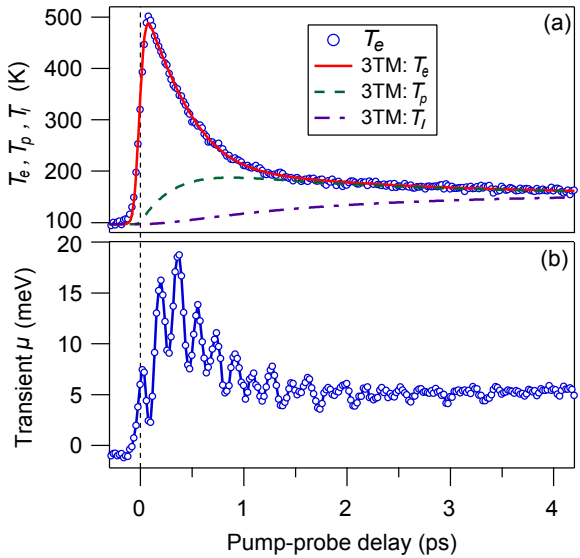}}%.eps
    \caption{\label{fig:f6 Te-Ef vs delay}\footnotesize{
    Time-dependent electronic temperature $T_e$ (a) and chemical potential $\mu$ (b) determined by fitting the transient trARPES spectra shown in figure \ref{fig:f1_spectrum+Ef} (see text). The results of a fit with a three temperature model (3TM) for $T_e$ (red), the temperature of a heat bath of more strongly coupled phonon modes $T_p$ (dashed green) and the lattice temperature $T_l$ (dashed-dotted dark violet) are also shown in (a).}}
\end{figure}

So far we have analyzed the oscillations in time-dependent photoelectron intensity. As it has been mentioned above, two further phenomena are
observed in the data presented in figure \ref{fig:f1_spectrum+Ef}. These are (i) the photo-excited distribution of electrons and holes and (ii) a
variation of the high-energy cutoff of the spectra. While the distributions of hot charge carriers have been discussed before already in detail for other
materials~\cite{Lisowski(2004), Perfetti(07), Fann1992, DelFatti2000, Rhie2003}
the clear change in the high energy cutoff is less established. Here we report the results of an analysis based on the assumption that these two phenomena can be described by the combination of a thermalized hot electron distribution function and a transient modification of the chemical potential. The fits presented in figure \ref{fig:f1_spectrum+Ef}(b) are the result of this model approach and describe the data very well. These fits have been performed as a function of time delay and the resulting hot electron temperature $T_e(t)$ and chemical potential $\mu(t)$ are depicted in figure \ref{fig:f6 Te-Ef vs delay}(a) and
\ref{fig:f6 Te-Ef vs delay}(b), respectively. Note that for this analysis a constant auxiliary density of states was assumed and non-thermal hot electrons which occur before the thermalization of the electronic system close to $t_0$ were neglected here for simplicity \footnote{For several systems we have observed and partly included such a subsystem of non-thermalized electrons \cite{Lisowski(2004), Perfetti(07)}.}.
We find a strong modulation of $\mu(t)$ which indicates that the intensity oscillations discussed above originate from a time-dependent change $\mu=\mu(t)$ and that the coherent phonon excitation modifies the
chemical potential. Although details of this phenomenon remain to be analyzed in future work we tentatively explain the changes in $\mu(t)$ as a consequence
of the modulation of the electronic structure in response to coherent phonons~\cite{Perfetti2006, Schmitt2008, Schmitt2011, Rettig2012a, loukakos_PRL07}. Due to the significant variation  of the electronic density of states near $E_F$ in the two band structure of FePns systems~\cite{Zbiri2009}, charge neutrality might well require a transient readjustment of $\mu$ to the instantaneous position of the electron and hole pocket at $X$ and $\Gamma$, respectively.

To describe the time-dependent change of $T_e$ we employ a three temperature model (3TM) as used before for the High $T_c$ superconductor $\text{Bi}_2\text{Sr}_2\text{CaCu}_2\text{O}_{8+\delta}$ (BSCCO)~\cite{Perfetti(07)}. This description gives a reasonable agreement with the obtained $T_e(t)$ as shown in figure \ref{fig:f6 Te-Ef vs delay}(a).  A similar analysis within the same 3TM has been recently performed for FePns using time-resolved optical reflectivity measurements~\cite{Mansart2010}. The 3TM describes the preferential coupling of the optically excited hot electron distribution with temperature $T_e$ to a more strongly coupled subset of phonon modes with temperature $T_p$, which subsequently couple on a longer timescale to the rest of the lattice modes with temperature $T_l$. The following set of coupled differential equations is used~\cite{Perfetti(07),Allen1987}:

\bea
\frac{\pd T_e}{\pd t} &=& -H(T_e, T_p) + \frac{S}{C_e}\\
\frac{\pd T_p}{\pd t} &=& +\frac{C_e}{C_p}H(T_e, T_p) - \frac{T_p-T_l}{\tau_\beta}\\
\frac{\pd T_l}{\pd t} &=& +\frac{C_p}{C_l}\frac{T_p-T_l}{\tau_\beta}
\eea

The source term $S$ is a Gaussian excitation pulse with a FWHM of $55\,\mathrm{fs}$ and an energy density determined by the absorbed fluence distributed over the optical penetration depth. The specific heat of electrons is determined by $C_e=\gamma_e T_e$, with the electronic specific heat coefficient $\gamma_e$. For the specific heat of the hot phonons and the rest of the phonons, the Einstein model with a mode $\omega_0$ is used, $C_p=3f \hbar\omega_0\left.\frac{\pd n}{\pd T}\right|_{T_p}$ and $C_l=3(1-f)\hbar\omega_0\left.\frac{\pd n}{\pd T}\right|_{T_l}$, where $n=(e^{\hbar\omega_0/k_B T} - 1)^{-1}$ is the Bose-Einstein distribution function. The parameter $f$ is the fraction of modes that are more strongly coupled to the electrons. The anharmonic decay (responsible for the loss of oscillation signal) of the respective subset of hot phonons to the remaining, more weakly coupled phonons, is described by $\tau_\beta$. For the energy transfer from the electrons to the more strongly coupled phonons, the formula derived by Allen~\cite{Allen1987} is used:

\be
H(T_e, T_p)=\frac{3\hbar\lambda\left\langle\omega^2\right\rangle}{\pi k_B}\frac{T_e-T_p}{T_e}\quad,
\ee

where $\lambda\left\langle\omega^2\right\rangle$ is the second moment of the Eliashberg e-ph coupling function $\alpha^2F(\omega)$. The e-ph coupling to the weakly coupled fraction $(1-f)$ barely influences the evolution of $T_e$ and has therefore been neglected. Electronic diffusion processes can be assumed not to be  essential here as the electric and thermal conductivity is strongly reduced along the longer c-axis due to the layered, quasi 2D structure of the FePns~\cite{Tanatar2009, Reid2010}.

Naturally, the description of the phonon system with a simple Einstein mode is a strong simplification and a proper choice of the mode energy influences the result of the simulation. To provide an estimate for the frequency $\omega_0$, we consider the phonon spectrum of the 122 FeAs compounds. The calculated phonon spectrum extends up to $\sim~35\,\mathrm{meV}$ as derived in several calculations~\cite{Boeri2010,Zbiri2009, Reznik2009} and supported by experiments~\cite{Zbiri2009, Hahn2009, Mittal2009} with the strongest contributions to $\alpha^2F(\omega)$ between $10-30\,\mathrm{meV}$~\cite{Boeri2010}.

The 3TM was fitted to the transient $T_e$ for various values of $\omega_0$ between $10-30\,\mathrm{meV}$. The resulting values for the parameters of the model are summarized in Table~\ref{tab:FeAs_3TM} for the three investigated compounds. For $f$ and $\lambda\left\langle\omega^2\right\rangle$, the left and right values are for a choice of $\omega_0=10\,\mathrm{meV}$ and $30\,\mathrm{meV}$, respectively. An exemplary fit of the electronic temperature $T_e$ using $\omega_0=18\,\mathrm{meV}$ is shown in figure~\ref{fig:f6 Te-Ef vs delay} (a) for the Co doped sample as solid line, which shows excellent agreement with the measured data. For other choices of $\omega_0$, similar good agreement to the data is found. The temperatures of hot phonons ($T_p$) and other phonons ($T_l$) derived from the model are shown as dashed and dash-dotted lines, respectively.

\begin{table}[htb]\begin{center}{
	\begin{tabular}{c||ccc}
	compound & $f$ & $\lambda\left\langle\omega^2\right\rangle$ ($\mathrm{meV^2}$) & $\tau_\beta$ ($\mathrm{ps}$)\tabularnewline
	\hline\hline
	EuFe$_2$As$_2$ & $0.43-0.38$ & $56-65$ & $2.9\pm0.3$ \tabularnewline
	BaFe$_{1.85}$Co$_{0.15}$As$_2$ & $0.48-0.43$ & $46-55$ & $3.3\pm0.5$ \tabularnewline
	BaFe$_2$As$_2$ & $\sim0.45$ & $30-46$ & $\sim3.5$ \tabularnewline
	\end{tabular}
	\caption{\footnotesize{Fitting parameters of the 3TM. Left and right values for $f$ and $\lambda\left\langle\omega^2\right\rangle$ correspond to a choice of $\omega_0=10\,\mathrm{meV}$ and $30\,\mathrm{meV}$, respectively.}}
	\label{tab:FeAs_3TM}
	}
\end{center}\end{table}

The values determined for $\lambda\left\langle\omega^2\right\rangle$ can be compared to recently published values for the e-ph coupling strength obtained from optical pump-probe  experiments. Mansart et al.~\cite{Mansart2010} report values of $\lambda\left\langle\omega^2\right\rangle\approx64\,\mathrm{meV^2}$ for almost optimally doped BaFe$_{2-x}$Co$_{x}$As$_2$, which was determined by the 3TM. Stojchevska et al.~\cite{Stojchevska2010} derived a somewhat higher value of $\lambda\left\langle\omega^2\right\rangle=110\pm10\,\mathrm{meV^2}$ for SrFe$_2$As$_2$ from the temperature dependence of quasiparticle relaxation times and for SmFeAsO$_{1-x}$F$_x$ an even larger value of $\lambda\left\langle\omega^2\right\rangle=135\pm10\,\mathrm{meV^2}$ is reported~\cite{Mer10}. In general, our results are compatible with the weak e-ph coupling found in the FePns and indicate even weaker coupling for BaFe$_2$As$_2$ compounds than in EuFe$_2$As$_2$. However, one has to bear in mind that the performed analysis assumes a thermalized electronic system and systematically neglects non-thermal electrons. As thermalization of non-thermal hot electrons leads to an additional heating of the thermalized electron distribution, this process can counteract the energy transfer to the lattice and leads to a slower energy relaxation of the thermalized electrons. Thus, the analysis of only the thermalized part of the electronic distribution might considerably underestimate e-ph coupling especially at early times.

The fraction $f\approx0.4$ of preferentially coupled modes is in agreement with the observations of Mansart et al.~\cite{Mansart2010}, and is considerably higher than found in BSCCO by $f\approx0.2$~\cite{Perfetti(07)}. Here we follow the explanation given by Mansart et al., who suggest that this stronger selectivity of e-ph coupling in the cuprates can be explained by the stronger 2D character of the cuprates, which could favor the selective coupling to specific modes. The FePns and especially Co doped 122 compounds are characterized by a stronger interlayer coupling and a more 3D character~\cite{Vilmercati2009}, which might lead to a less selective e-ph coupling~\cite{Mansart2010}. The equilibration time $\tau_\beta$ of the hot phonon distribution with the rest of the lattice modes is considerably faster here than in the optical reflectivity measurements by Mansart et al., who report $\tau_\beta\sim5-7\,\mathrm{ps}$. This faster relaxation of excited phonon modes observed in trARPES measurements compared to optical reflectivity is consistent with the faster relaxation of the coherently excited phonon modes discussed earlier and indicates a different phonon relaxation behavior in the bulk and the surface/subsurface region, respectively. In addition, the excitation densities in the experiments by Mansart et al. where a factor of two to three larger than here, and an increase of $\tau_\beta$ with fluence was observed~\cite{Mansart2010}.

Based on our results of $\lambda\left\langle\omega^2\right\rangle$ we can estimate the value of the e-ph coupling constant $\lambda$ for a particular value of $\omega$. Considering the coherently excited A$_{1g}$ mode at $23\,\mathrm{meV}$, which showed enhanced coupling in calculations, we find $\lambda<0.15$ for all compounds. This estimate is in agreement with calculations~\cite{Boeri1(2008), Boeri2010} of various FePns compounds, which report average values of $\lambda<0.35$. Taking the mean of the phonon DOS as reference, $\lambda$ gets even smaller in agreement with other publications. Even if we consider the lowest coupled modes around $12\,\mathrm{meV}$ to be most important for e-ph coupling, $\lambda$ does not exceed a value of 0.5. Similar small values for $\lambda$ have been found in the cuprates~\cite{Perfetti(07)}, which suggests limited importance of e-ph coupling for the pairing mechanism in both classes of materials.

%##################################################################################################

%****** SUMMARY
\section{Conclusion}
Femtosecond time- and angle-resolved photoelectron spectroscopy was demonstrated to
be a powerful tool to investigate the response of the electronic system in the
vicinity of the Fermi level. In this study we could separate for various 122
FePns compounds transient changes in the electron distribution function and
the chemical potential. The first one was analyzed by a simplified three temperature
model. In agreement with earlier
time-resolved optical studies we conclude a rather small momentum-averaged electron
phonon coupling constant $\lambda<0.15$, which suggests a limited importance of e-ph
coupling for the Cooper pair formation in these systems.
The transient chemical potential is explained to be the consequence of charge
neutrality in the probed region of the material while the electronic system is in a
non-equilibrium state characterized by hot electrons and coherent as well as
incoherent phonons. Such changes in the chemical potential can be expected to be a
general phenomena, however, we consider that the two band nature of the 122
FePns compounds enhances the effect in comparison to a single band system. The
pronounced effect of coherent phonons on the electronic structure near $E_F$
demonstrate vivid electron-phonon coupling but the assignment of the two
observed lower frequency excitations beside the A$_{1g}$ mode requires further
attention. Future studies as function of doping might provide more insight into the
origin of the changes in the chemical potential as well as the material specificity
of all observed coherent phonon modes.
%##################################################################################################

%**** ACKNOWLEDGEMENTS
\section*{ACKNOWLEDGEMENTS}

Funding from the Deutsche Forschungsgemeinschaft within SPP 1458 is gratefully acknowledged. R.C Acknowledges the Alexander von Humboldt Foundation.
We thank Hermann D\"{u}rr and Lilia Boeri for fruitful discussions.
%################################################################
\bibliography{Fe_Pnictides}%.bib

\end{document}